\documentclass[12pt,preprint]{aastex}
\usepackage{graphicx}
\def\kms{km\,s$^{-1}$}
\def\ngc{NGC7538-IRS1\,N\,}
\def\Mo     {\hbox{$\rm M_{\odot}$}}
\def\eq#1{\begin{equation} #1 \end{equation}}
\def\about  {\hbox{$\sim$}}
\def\half   {\hbox{$\frac12$}}
\def\Dv     {\hbox{$\Delta v_{\rm D}$}}

\def\vk     {\hbox{$v_{\rm k}$}}
\def\vmin   {\hbox{$v_{\rm min}$}}
\def\vmax   {\hbox{$v_{\rm max}$}}
\def\Ri     {\hbox{$R_{\rm i}$}}
\def\Ro     {\hbox{$R_{\rm o}$}}
\def\ri     {\hbox{$\rho_{\rm i}$}}
\def\ro     {\hbox{$\rho_{\rm o}$}}
\def\Oi     {\hbox{$\Omega_{\rm i}$}}
\def\Oo     {\hbox{$\Omega_{\rm o}$}}
\def\Om     {\hbox{$\Omega_{\rm m}$}}
\def\Obar   {\hbox{$\bar \Omega$}}
\def\tetak  {\hbox{$\theta_{\rm k}$}}
\def\tauo   {\hbox{$\tau_0$}}

\def\ltsimeq{\,\raise 0.3 ex\hbox{$ < $}\kern -0.75 em
 \lower 0.7 ex\hbox{$\sim$}\,}
\def\gtsimeq{\,\raise 0.3 ex\hbox{$ > $}\kern -0.75 em
 \lower 0.7 ex\hbox{$\sim$}\,}

\begin{document}

\title{A circumstellar disc in a high-mass star forming region}

\shorttitle{Disc in NGC7538} \shortauthors{Pestalozzi et al.}

%\slugcomment{ME revisions; January 19, 2004}

\author{Michele R. Pestalozzi\altaffilmark{1},
        Moshe Elitzur\altaffilmark{2},
        John E. Conway\altaffilmark{1}
        and Roy S. Booth\altaffilmark{1}}

\altaffiltext{1}{Onsala Space Observatory, SE-43992 Onsala, Sweden}
\altaffiltext{2}{Department of Physics and Astronomy, University of Kentucky,
           Lexington, KY 40506--0055, USA}

\email{michele@oso.chalmers.se, moshe@uky.edu,
       jconway@oso.chalmers.se, roy@oso.chalmers.se}

\begin{abstract}

We present an edge-on Keplerian disc model to explain the main component of the
12.2 and  6.7\,GHz methanol maser emission detected toward \ngc. The brightness
distribution and spectrum of the line of bright masers are successfully modeled
with high amplification of background radio continuum emission along velocity
coherent paths through a maser disc. The bend seen in the
position-velocity diagram is a characteristic signature of differentially
rotating discs. For a central mass of 30\Mo, suggested by other observations,
our model fixes the masing disc to have inner and outer radii of $\sim350$\,AU
and $\sim1000$\,AU.
\end{abstract}

\keywords{star formation -- masers -- Interstellar medium -- discs}

\section{Introduction}

Disks are expected to form during protostellar collapse, and low-mass stars
seem to provide good observational evidence for the existence of disks (e.g.
\citealt{qi03,fue03}). The situation is less clear for high-mass stars. While
in several cases velocity gradients in massive star forming regions have been
detected on large scales ($>10,000$AU, e.g. \citealt{san03}), evidence for
compact discs on scales $\la 1000$\,AU remains sparse. One possible example is
IRAS20126+4104, where there is good evidence for disc-outflow geometry around a
24\Mo\ central object \citep{ces99,mol00}. The case for a disc around
G\,192.16-3.82, a protostar of \about 15\Mo\ \citep{she01}, is less compelling.

Class II methanol maser emission, a signpost of high mass star formation
\citep{min01}, offers a potential indicator of discs since it often shows
linear structures in both space and position --- line-of-sight velocity
diagrams \citep{nor98,min00a}. These masers have been modeled as occurring at
fixed radii within edge-on discs, although it is still unclear whether they
arise primarily in discs or outflows \citep{deb03,mos02}. One of the most
striking examples of a maser line in both space and velocity is
found in \ngc \citep{min98,min00a}. We present here the first quantitative
Keplerian disc analysis of this maser without invoking the assumption of a
single radius. We find compelling evidence for the disc interpretation in this
case.

\begin{figure*}[!ht]
  \begin{center}
    \includegraphics[width=0.48\hsize]{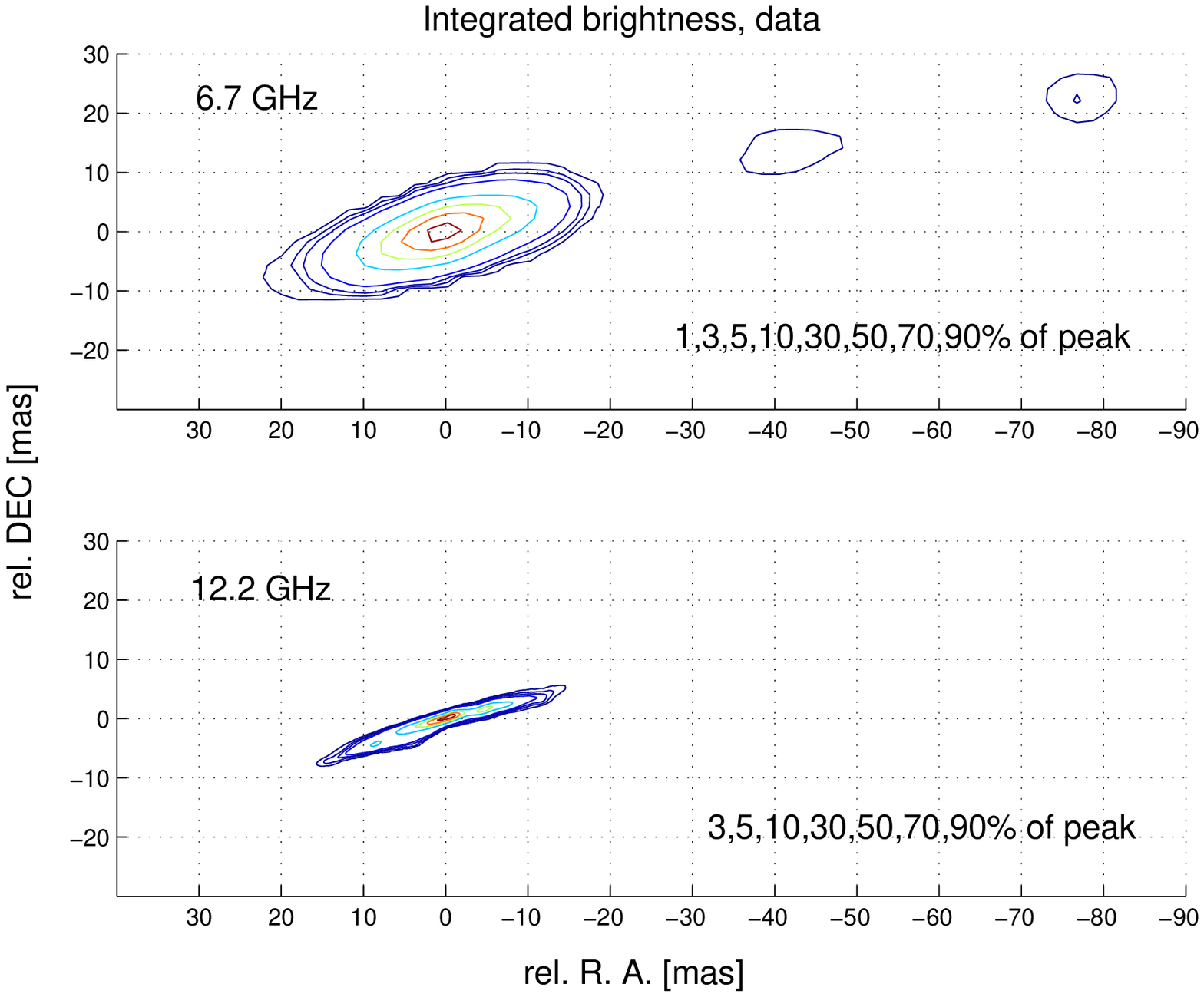} \hfil
    \includegraphics[width=0.48\hsize]{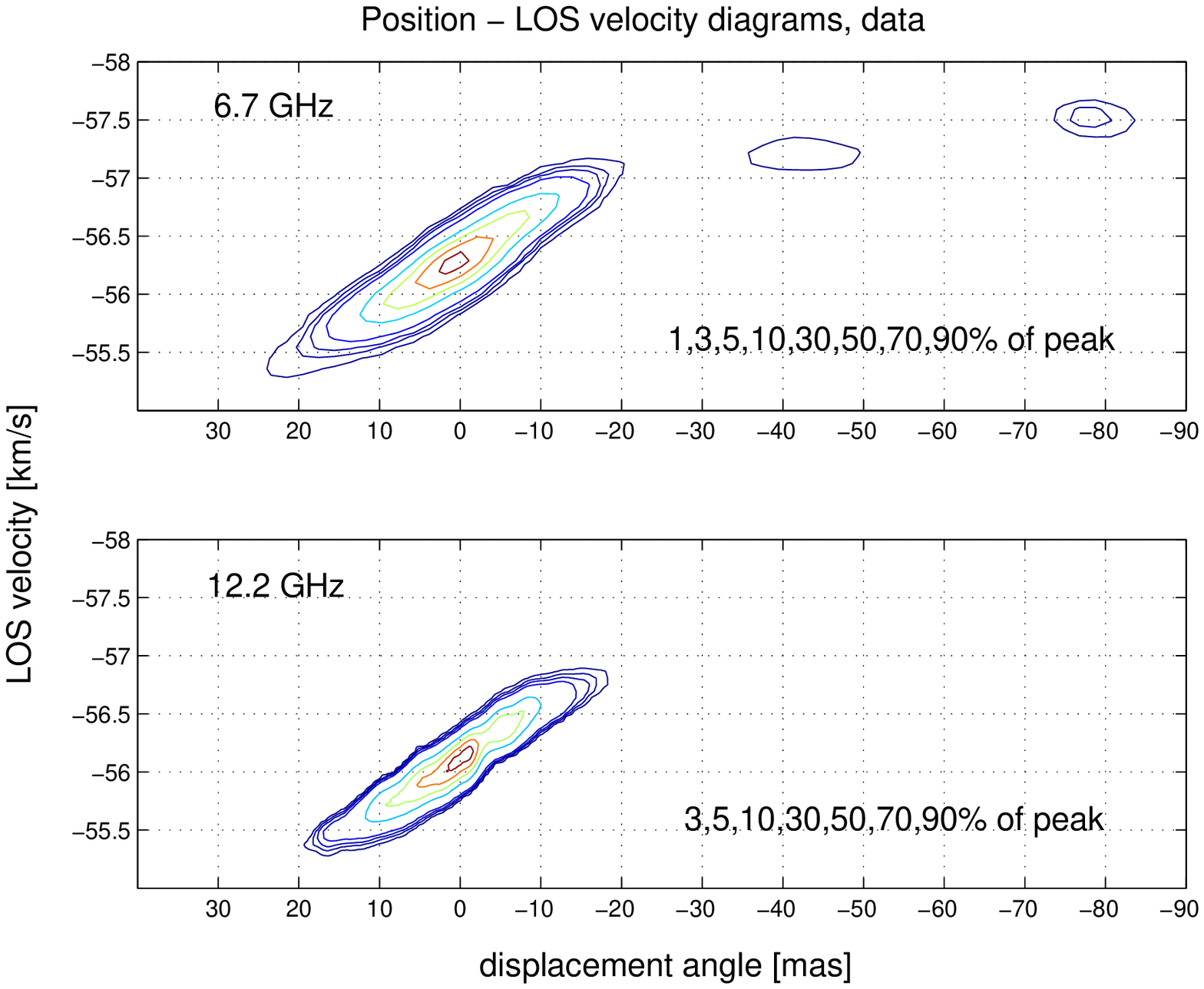}
    \caption{Left: Zero moment (integrated emission over velocity) maps of
    the 6.7 and 12.2\,GHz methanol masers in  the main spectral feature of
      \ngc. Right: ($\theta,v$)-diagrams of the
      two masers, in which the data have been averaged over the spatial
      direction perpendicular to the major axis position angle. }
    \label{fig:data_all}
  \end{center}
\end{figure*}

\section{Maser data and other observations}

The NGC7538-IRS1 star formation region has been extensively studied at several
wavelengths. The spectral type of the central star has been estimated as O6
(\citealt{wil76,cam84b}), in agreement with the SED from 2\,mm
(\citealt{aka01}) to 850\,$\mu$m (\citealt{mom01}) which implies a central
object with luminosity of $>8.3 \times 10^{4} L_{\odot}$ and mass of about
30\Mo. VLA observations uncovered an associated ultra-compact HII region with
peak brightness temperature 10,000--15,000\,K (\citealt{cam84,gau95}). Radio
free-free continuum and recombination line observations show a N-S elongated
structure attributed to a large scale outflow.

The first VLBI observations of the 12.2\,GHz methanol masers
(\citealt{min98,min00a}) revealed several groups of spots corresponding to
different features in the single dish spectra. The subject of our study is the
maser in the main spectral feature between about $-56.50$ and $-56.82$ \kms.
The 12.2\,GHz data cube from published data by \citet{min01}, made from 1998
VLBA\footnote{The VLBA is operated from the National Radio AStronomy
  Observatory's Array Operation Center in Socorro, NM.} observations, is shown
in Figure \ref{fig:data_all}. The interpretation of 
the linear structure in space and velocity as an edge-on disc (\citealt{min98})
is supported by the positional agreement of the maser line with the waist of
the hour-glass shaped continuum radio emission (\citealt{gau95}) and a band of
free-free absorption. The same structure is evident in the 6.7\,GHz maser data,
obtained in February 2001 with the European VLBI Network (EVN\footnote{
The EVN is a joint facility of European, Chinese, South African and other
radio astronomy institutes.}), shown in the 
figure. Taking into account 
the different angular resolution, the velocity-integrated emission of the two
masers is similar out to $\pm$20\,mas in R.A. (Figure \ref{fig:data_all},
left). In addition, the 6.7\,GHz data shows two disconnected `outlier' maser
features which lie close to the extrapolation of the maser line. The right
panels in Figure \ref{fig:data_all} show the position-velocity diagrams for the
two masers. They both show the same velocity gradient near the center and a
bend at $\sim$15\,mas.

The 12.2\,GHz line involves the $2_0 \to 3_{-1}E$ transition, the 6.7\,GHz line
the $5_1 \to 6_0A^+$ transition. Since $E$- and $A$-methanol can be regarded as
distinct molecules for all practical purposes, the two masers can be expected
to probe different physical conditions even when they reside in the same
region. Yet in spite of these differences, the brightness peaks of the two
lines are perfectly coincident in both velocity and spatial location. The
agreement is within the 12.2\,GHz angular resolution of 2--3 mas and the
spectral resolutions of 0.048 and 0.09\,\kms per channel for the 12.2- and
6.7\,GHz data, respectively. Convolving the 12.2\,GHz data to the 6.7\,GHz
spatial resolution we find that the ratio of the two emissions is constant to
within 10\% over the entire feature.

\section{Modeling}

The remarkable coincidence in space and frequency of the two masers is in fact
a natural consequence of amplification of a background source by an edge-on
rotating disk. In that case the maser brightness at displacements $\theta$ and
$v$ from the position and velocity of the line-of-sight (LOS) diameter is
$I(\theta,v) = I_{\rm B}\,e^{\tau(\theta,v)}$, where $I_{\rm B}$ is the
background continuum and $\tau(\theta,v)$ the maser (negative) optical depth.
Irrespective of the disk annuli where the masers reside, the peak emission will
always occur at $(\theta,v) = (0,0)$, which picks out the largest optical depth
\tauo. With brightness temperature of $10^4$~K for the background continuum,
the function $\tau(\theta,v) = \ln\left[I(\theta,v)/I_{\rm B}\right]$ is a
measured quantity. We find $\tau_0 = 18.32$ and 15.99 for the 6.7- and
12.2\,GHz masers, respectively.

Consider an edge-on rotating disc at distance $D$ and a point at radius $\rho =
r/D$ along a path with displacement $\theta$. The rotation velocity $V(\rho)$
and its LOS component $v$ obey $v/\theta = V/\rho = \Omega$, the angular
velocity. In Keplerian rotation $\Omega \propto \rho^{-3/2}$ with $\Oo =
\Omega(\ro) = D \sqrt{GM/\!R_{\rm o}^3}$ where $\ro = \Ro/D$ is the outer
radius and $M$ the central mass. Assume a Gaussian frequency profile for the
maser absorption coefficient and denote the radial variation of its magnitude
at line center by the normalized profile $\eta(\rho)$ ($\int\eta\,d\rho = 1$).
Then
\eq{\label{eq:tau}
 \tau(\theta,v) =  \tauo \int\!\eta(\rho)
      \exp\left[-\half\!\left(v -  \Omega(\rho)\,\theta\over\Dv\right)^{\!2}
      \right] {d\rho\over\beta}
}
where $\beta = \sqrt{1 - (\theta/\rho)^2}$ and $d\rho/\beta$ is distance along
the path. The width \Dv\ is taken as constant, although it could vary with
$\rho$ because of temperature variation and maser saturation. Saturation
typically requires $\tau \ga$ 10--15 across the disc radius, and the broadening
of the absorption profile is then proportional to length in excess of this
threshold (\citealt{eli92c}). If the \ngc disc is located in front of the
background source then the overall amplification with $\tau$ \about\ 18 implies
an optical depth of only 9 across the radius, too low for saturation. If the
radio continuum is centered on the star then the optical depth across the
radius is \about\ 18 and saturation can be expected to have an effect, but only
at the very outer segments of the disc where a decrease in temperature could
have an opposite, offsetting effect. We will consider these effects in a future
publication.

\begin{figure}[ht!]
  \begin{center}
    \includegraphics[width=\hsize]{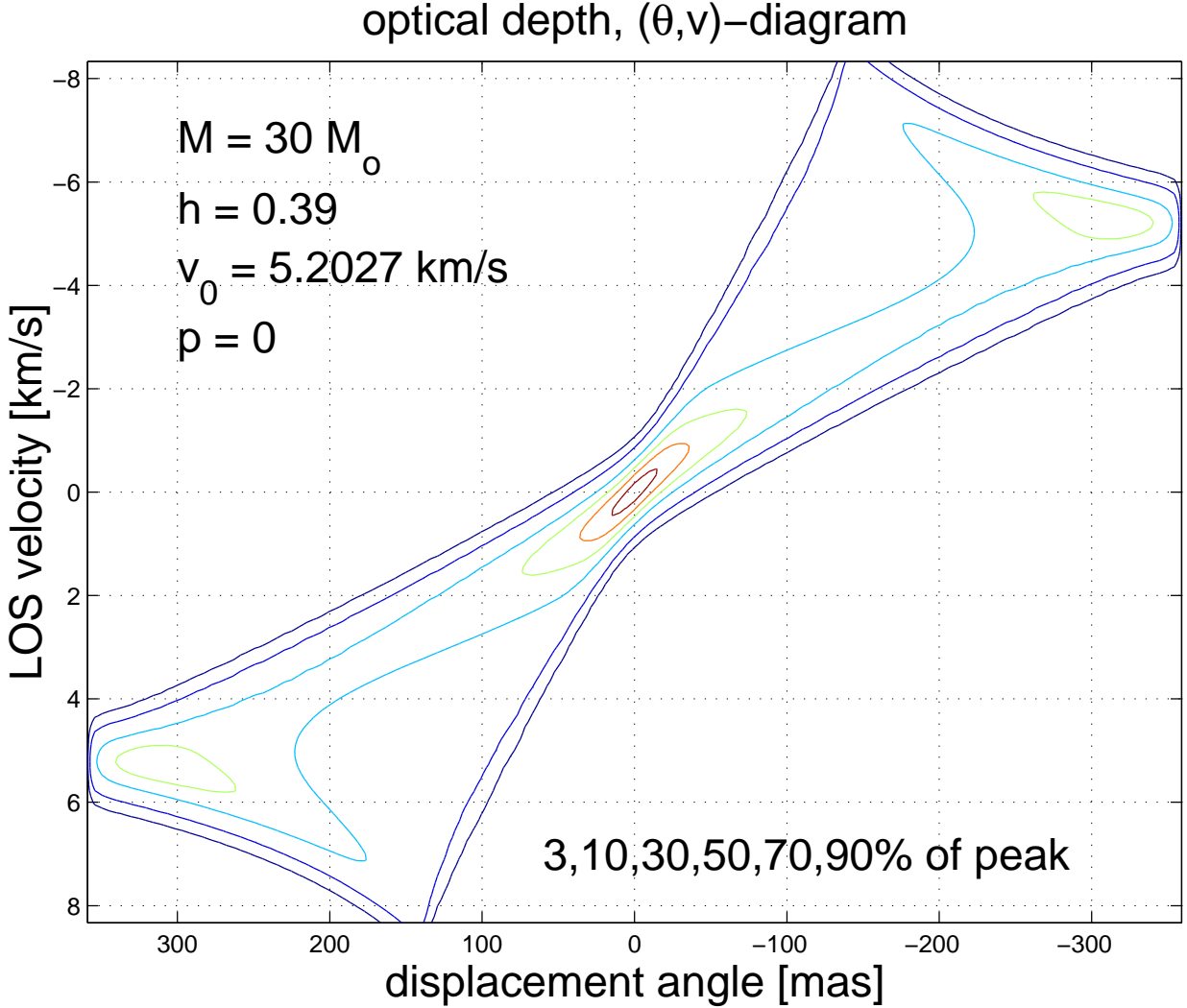}
\caption{Contour plots of $\tau(\theta,v)$ (see eq.\ \ref{eq:tau}) for a
Keplerian disc with \Dv = 0.40 \kms\ around a 30\Mo\ star, viewed edge-on
from $D$ = 2.7 kpc. Constant maser absorption coefficient between radii \Ri\ =
378\,AU (0\farcs14) and \Ro\ = 1080\,AU (0\farcs4).}
\label{fig:bv_tot}
  \end{center}
\end{figure}

We parameterize $\eta$ with a power law $\rho^{-p}$ from an inner radius $\ri =
\Ri/D$ to \ro. Both \tauo\ and \Dv\ are set directly from $\tau(0,v)$. Modeling
the full $\tau(\theta,v)$ requires two free parameters for the integrand (\Oo\
and $p$) and two for the integration limits (\ro\ and $h = \ri/\ro$). Figure
\ref{fig:bv_tot} shows contour plots of $\tau(\theta,v)$ for a representative 
Keplerian disc around a 30\,\Mo\ protostar with the \Dv\ determined from the
12.2 GHz data at $\theta=0$. The system is viewed edge-on at the adopted
distance to NGC7538.

The lowest contour (3\%) outlines the angular velocity $\Omega = v/\theta$ on
the disk boundaries. The shallow segments of this contour (lower branch for
$\theta < 0$) trace out the disk outer radius. The steeper ones trace the inner
radius when $|\theta| < \ri$ and the mid-line when $|\theta| > \ri$, where the
Keplerian $v \propto |\theta|^{-1/2}$ is evident. Along any path at $|\theta| <
\ri$, the velocity varies from \vmin\ = $\Oo|\theta|$ on \Ro\ to \vmax\ =
$\Oi|\theta|$ on \Ri. As long as $\vmax - \vmin < \Dv$, the full path remains
velocity coherent and the maximal $\tau$ is centered on the velocity determined
from $\partial\tau(v,\theta)/\partial v = 0$. This yields $v = \Obar\,\theta$
where $\Obar = \int \Omega\,\eta d\rho$, an average dominated by the high
angular velocity of the inner regions; for example, constant $\eta$ gives
$\Obar = 2\Oi h/(1 + h^{1/2})$. The linear relation between $v$ and $\theta$ is
evident in the uniform inclination of the innermost contours.

The condition $\vmax - \vmin = \Dv$ is met at a displacement \tetak, and $\tau$
gets contributions from only a fraction of the path when $\theta > \tetak$.
Since the longest coherence is at the outer radius, where $\Omega$ is smallest,
the $\tau$-contours bend toward lower velocities as is evident in the figure
around $\tetak \approx 0\farcs02$. This bending reflects the change from
inner-edge to outer-edge dominance of longest coherence.
As $\theta$ increases further beyond \ri, the longest coherent segment moves
from the outer radius to the disc mid-line. The maximal $\tau$ begins to
increase, creating the local outer maximum evident in the 50\% contours at
$\theta \approx 0\farcs39$ before final truncation at $\theta = \ro$.

This discussion shows that the contours of highest amplification, down to 70\%
of $\tau_0$, contain dependence on $p$, $h$ and $\Omega$ but not on \ro. This
has two important consequences: (1) Successful modeling of the strongest maser
emission requires one less parameter. (2) Since $\Omega$ involves only the
combination $M/R^3$, the central mass cannot be determined without observations
of (exponentially weaker) radiation close to the phase-space boundary to
determine \ro.

\begin{figure}[h]
  \begin{center}
    \includegraphics[width=\hsize]{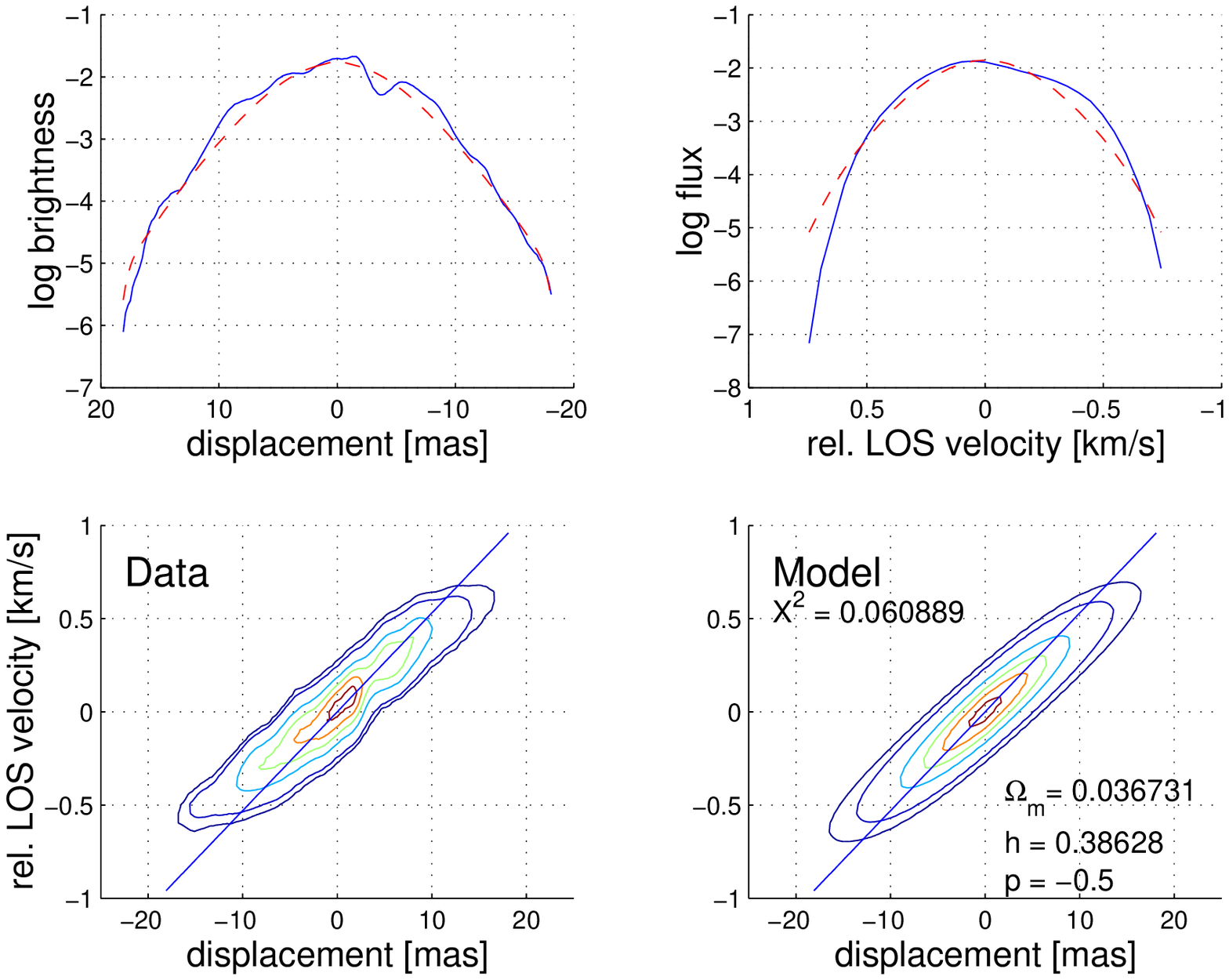}
\caption{Modeling the 12.2\,GHz data with a disc with the indicated parameters
  and \Dv = 0.4 \kms.
Top: $\theta$-variation of frequency-integrated brightness (left) and
$v$-variation of the flux (right). In both diagrams the log of the quantities
is shown. Solid lines show the data, dashed lines the model. Bottom: Contour
plots of $I(\theta,v)$, the data to the left, the model to the right. The
contours of the model have been obtained from the exponential of eq.
\ref{eq:tau}. Contours are 5,10,30,50,70,90\% of peak. The slope of the
diagonal line is \Oi\,in \kms/mas. }
    \label{fig:fit}
  \end{center}
\end{figure}

Brightness measurements with dynamic range $f$ map the $\tau$-distribution
between $\tau_0$ and $\tau_0 - \ln f$. With $\tau_0 $ = 15.99 and $f \simeq
100$, the 12.2\,GHz data traces the $\tau$-contours in \ngc down to 70\% of
peak. Therefore the data shown in the lower left map of figure \ref{fig:fit}
cover only the two innermost contours in figure \ref{fig:bv_tot}, and if our
model is applicable it should require only the three parameters $h$, $\Omega$
and $p$. In fact, the first two can be determined quite accurately even without
detailed modeling since the bend in the $\tau$-contours occurs at $\vk \simeq
\half\Dv\left(1 + h^{3/2}\right)\!/\!\left(1 - h^{3/2}\right)$ and $\tetak
\simeq \Om\vk$, where $\Om = \half(\Oi + \Oo)$. We find directly that $h$ =
0.39 and \Om\ = 0.037\,\kms/mas, implying $M/R_{\rm o}^3 = 30\Mo/(1000\rm
AU)^3$ and leaving $p$ as the only free parameter to match all other data
points in the $(\theta,v)$-plane. Sensible fits to the 12.2\,GHz data are
obtained when $0.035 \ltsimeq \Om \ltsimeq 0.045$ and $0.3 \ltsimeq h \ltsimeq
0.4$.

The lower right panel of figure \ref{fig:fit} shows the model map for $p =
-0.5$, the top panels show how well the fitting of $\tau(\theta,v)$ reproduces
the brightness profile and the spectrum. The negative $p$ means that $\eta$,
i.e., the maser absorption coefficient, increases with radius. The relation
between $\eta$ and overall density involves the product of the methanol
abundance, the fraction of methanol in the maser system and the fractional
inversion. Any of these factors might give more weight to the disk outer
regions. Considering the simple power law with sharp cutoffs we employ, a 70\%
increase in $\eta$ across the disk is compatible with a constant density
profile.

\begin{figure}[h]
  \begin{center}
  \includegraphics[width=\hsize]{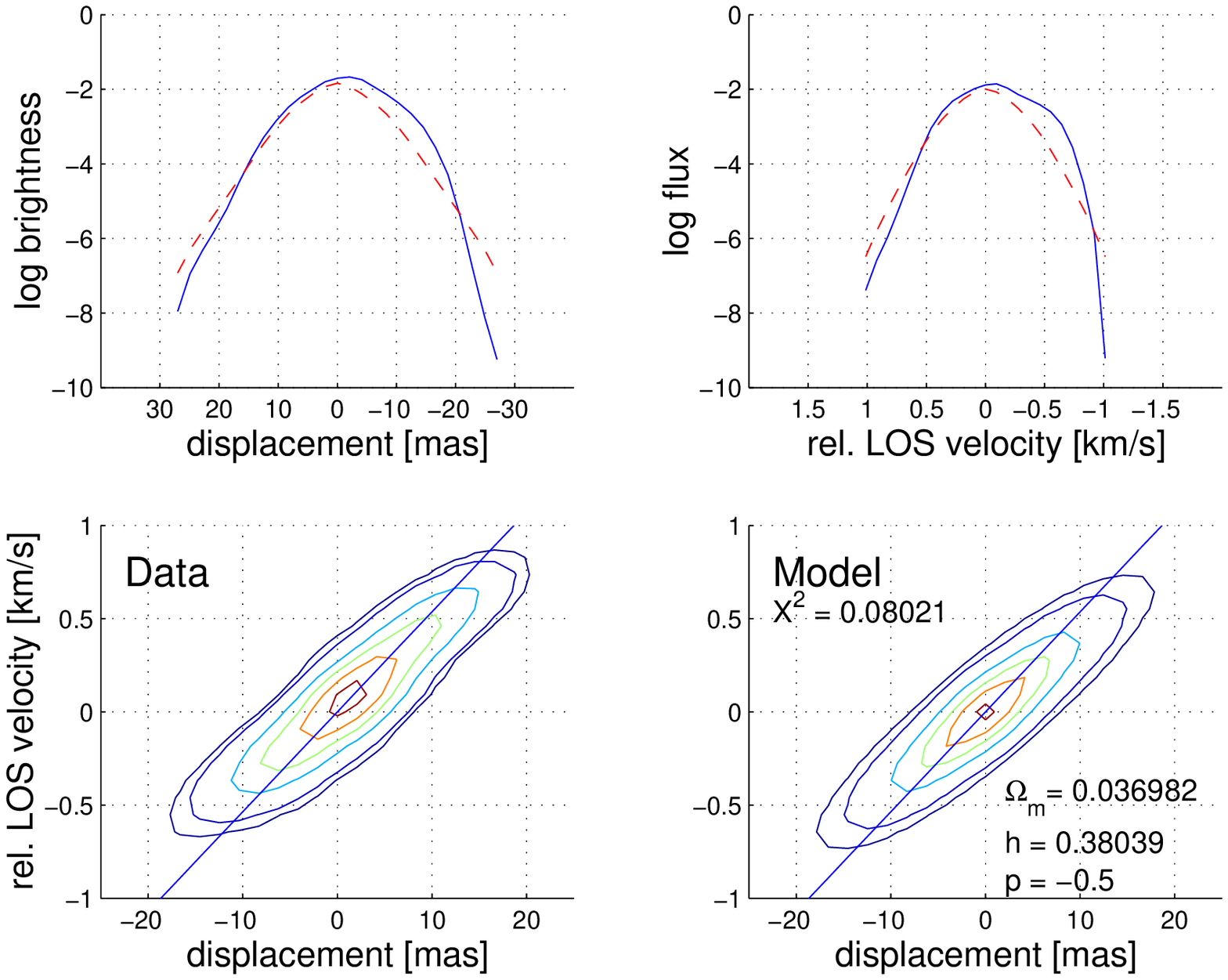}
\caption{Same as figure \ref{fig:fit} for the 6.7\,GHz maser.}
  \label{fig:fit_6.7}
  \end{center}
\end{figure}

Our model is equally successful in fitting the main 6.7\,GHz maser emission, as
is evident from figure \ref{fig:fit_6.7}.  The value of $p$ is the same for
this maser, the only slightly different parameters are \tauo\ = 18.32,
$h$=0.38 and \Om\ = 0.037 km/s/mas. The value for \Dv was taken from the
12.2\,GHz maser, which is considered to be more reliable because of the higher
spatial resolution. Good fits of the 6.7\,GHz data are obtained when $0.035
\ltsimeq \Om \ltsimeq 0.045$ and $ 0.35 \ltsimeq h \ltsimeq 0.45$. The virtual
identity of the disc segments of such different masing transitions is an
important challenge for class II methanol pumping calculations.

\section{Discussion}
\label{sec:dis_con}

The analysis presented here shows that a Keplerian disk explains successfully,
down to minute details, the main methanol maser feature in \ngc. The angular
velocity we find implies a disk outer radius of \about1000\,AU if the central
mass is 30\Mo, as determined by \cite{cam84}. The methanol maser emission
covers a substantial area of the disk, with the radius varying by a factor of
3.

The $\tau(\theta,v)$-contours display a bend in the position-velocity diagram.
Since $v = \Omega\,\theta$, a straight line in this diagram implies a constant
$\Omega(\rho)$, i.e., either a single radius or, in the case of an extended
region, solid-body rotation. Our analysis therefore uncovers a unique signature
of differential rotation in the \ngc  disc. The emission's smooth structure and
the observation's high sensitivity are crucial for this detection, which is
made well inside the \about\ 1.2 \kms\ width of the spectral feature (see top
right panels in figures \ref{fig:fit} and \ref{fig:fit_6.7}). These
requirements preclude detection of the effect in extragalactic sources and
explain why it could not be considered in the analysis of the H$_2$O maser disc
in NGC4258 (e.g., \citealt{wat94}). However, a bend in the position-velocity
diagram has been found in thermal line emission from low-mass stars and
interpreted as steep-gradient solid-body rotation inside a shallow-gradient
Keplerian outer region (\citealt{bec93, oha97,bel02}). Since maser and
optically thin thermal emission, which is directly proportional to optical
depth, share the same kinematic structure, our analysis shows there is no need
to invoke additional components on top of Keplerian rotation.

Our analysis is valid for a radio continuum source placed either behind the
maser disc or at its center. The latter possibility would fit well in the
context of disc photoevaporation models (\citealt{hol94}). In that case, the
radial extent of disc photoevaporation by the central star in NGC7538 covers
the entire maser region and could activate both methanol formation and the
observed outflow, responsible for the radio continuum. The maser location at
the outflow waistline makes this scenario especially attractive.

The methanol maser data determines detailed disc properties with high precision
yet the central mass remains undetermined since it requires the outer radius.
And \ro\ is still unobservable because the angular extent of the detected
emission does not trace the full disk, instead it is controlled by velocity
coherence and the dynamic range of the observations. The amplification opacity
at the disk outer edge (see fig.\ \ref{fig:bv_tot}) is 50\% of the central peak
level. With $\tau$ = 16 (18) at the center, detection of the outer peak
requires a dynamic range of 3000 (8100) if the radio source is located behind
the maser disc, placing it beyond the reach of the current observations by a
factor of \about\ 30 (80).  The requirements become even more severe if the
continuum is centered on the star. The 6.7\,GHz outlier features therefore
cannot correspond to the outer peak of our model disc. Instead, these outliers
are probably the first 
detection of enhanced amplification by small inhomogeneities in an otherwise
rather smooth disc. Such non-axisymmetric distortions of the disk surface, and
their possible effect on the pumping, have been proposed by \citet{dur01}. It
should be possible, though, to determine \ro\ by probing the boundary of the
($\theta,v$) region at $\theta < \ro$ along the $v$-axis or another direction
such as the normal to \Obar. We will study these issues in detail in a future
publication.

This study provides one of the few pieces of evidence to date for a compact
($\la 1000$ AU) disc around a massive protostar. It seems certain, though,
that not all class II methanol masers occur in discs. Even in \ngc itself, the
weaker methanol masers to the south, which are associated with other maser
species (OH, H$_{2}$O etc), are likely to arise from an outflow. The
interpretation of ordered lines of methanol masers in terms of disc or outflow
models must be decided case by case with detailed modeling.

\vskip -3ex
\acknowledgments
M. Pestalozzi thanks his fellow PhD student Rodrigo Parra for the useful
discussions and for all the useful MATLAB
tricks. M. Elitzur thanks the NSF for its support and Chalmers University for a
Jubileum Professor award that triggered a most enjoyable visit to
Onsala. The reduced 12.2\,GHz data cube was kindly made available by Vincent
Minier.

%The National Radio Astronomy Observatory is a facility of the
%National Science Foundation, operated under agreement by Associated
%Universities Inc.

%\bibliography{methanol,stars,othermasers,surveys_cat,tech,starformation+IR,varia,theory}
%\bibliographystyle{aa}

\end{document}